\documentclass[prb,showpacs,preprint,nobibnotes,nofootinbib]{revtex4}
\usepackage{graphicx}
\usepackage{multirow}
\begin{document}
\bibliographystyle{apsrev}

\title{Using \textsc{ONETEP} for accurate and efficient $\mathcal{O}(N)$ density
functional  calculations}

\author{Chris-Kriton Skylaris}
\thanks{Corresponding author}
\affiliation{Physical {\&} Theoretical Chemistry Laboratory, South
Parks Road, Oxford OX1 3QZ, UK}
 \email{chris-kriton.skylaris@chem.ox.ac.uk}
\homepage{http://www.chem.ox.ac.uk/researchguide/ckskylaris.html}
\author{Peter D. Haynes}
\author{Arash A. Mostofi}
\author{Mike C. Payne}
\affiliation{Theory of Condensed Matter, Cavendish Laboratory, \\
J. J. Thomson Avenue, Cambridge CB3 0HE, UK }

\date{\today}

\begin{abstract}
We present a detailed comparison between \textsc{onetep}, our
linear-scaling density functional method, and the conventional
pseudopotential plane wave approach in order to demonstrate its high
accuracy. Further comparison with all-electron calculations shows
that only the largest available Gaussian basis sets can match the
accuracy of routine \textsc{onetep} calculations. Results indicate
that our minimisation procedure is not ill-conditioned and that
convergence to self-consistency is achieved efficiently. Finally we
present calculations with \textsc{onetep}, on systems of about 1000
atoms, of electronic, structural and chemical properties of a wide
variety of materials such as metallic and semiconducting carbon
nanotubes, crystalline silicon and a protein complex.
\end{abstract}


\maketitle

\section{Introduction}

The formalism of Kohn-Sham density functional theory (DFT)
\cite{HOHENBERG1, KOHNSHAM} for electronic structure calculations
has become established as an approach that provides a good
description of electronic correlation while keeping the size of
calculations tractable. Nevertheless, the computational time taken
by a conventional DFT calculation increases with the cube of the
number of atoms. This scaling limits the size of problems that can
be tackled to a few hundred atoms at most. As a consequence, many
exciting problems which lie at the interface between the microscopic
and mesoscopic worlds, particularly in the fields of biophysics and
nanoscience, are out of the reach of DFT calculations. Progress
towards the goal of bringing the predictive power of DFT to bear on
these problems can be made only by developing approaches for DFT
calculations that have linear-scaling or $\mathcal{O}(N)$ instead of
cubic-scaling computational cost.

Even though there have been numerous theoretical developments, so
far linear-scaling methods have not lived up to their early promise.
Linear-scaling approaches are still described as ``experimental"
\cite{MARTIN2004} and so far there are few examples of successful
application to problems of interest in materials or biological
sciences \cite{ARTACHO2003}. For a review see
Refs.~\onlinecite{GALLI1996,GOEDECKER-REVIEW}. Our \textsc{onetep}
linear-scaling method for DFT calculations allows for the systematic
control of both truncation errors and variational freedom in the
basis set. For full details see Ref.~\onlinecite{SKYLARIS2005} and
references therein. Here we demonstrate that \textsc{onetep} can be
used to solve real problems with the same level of confidence and
general applicability as conventional cubic-scaling DFT approaches.

In section \ref{background-section} we begin
with a brief presentation of the formalism for
linear-scaling DFT on which \textsc{onetep} is based.
In section
\ref{convergence-section} we compare \textsc{onetep} with
conventional well-established cubic-scaling methods with emphasis on
the case of systematic improvement in the basis set, and hence in
accuracy, and in the speed of self-consistent convergence. In section
\ref{results-section} we show how \textsc{onetep} can be used to
explore a range of materials with thousands of atoms ranging from
nanostructures to bulk solids to biomolecules. Finally, in section
\ref{conclusions-section} we present our conclusions.

\section{Overview of theory} \label{background-section}

Kohn-Sham DFT enables the problem of many interacting electrons in a
static external potential to be mapped onto a fictitious system of
non-interacting particles. Self-consistent solution of the resulting
set of single-particle Schr\"{o}dinger equations gives the
ground-state energy and density of the original interacting problem.
All the information about the ground state of the system is
contained in the single-particle density matrix $\rho(\mathbf{r},
\mathbf{r}')$ which, provided there is a band gap in the material,
decays exponentially \cite{KOHN1959,DESCLOIZEAUX1964,
BAER1997,ISMAIL-BEIGI1999, HE2001} as a function of the distance
between $\mathbf{r}'$ and $\mathbf{r}$. This property can be
exploited to truncate the density matrix so that the amount of
information it contains increases only linearly with the number of
atoms. To perform this truncation in a practical way, the density
matrix is expressed as
\begin{equation}
\rho(\mathbf{r}, \mathbf{r}') =  \sum_{\alpha\beta} \phi_{\alpha}(\mathbf{r})
K^{\alpha \beta} \phi_{\beta}^{*}(\mathbf{r}')
\end{equation}
where the $\{ \phi_{\alpha}\}$ are a set of spatially
localised, nonorthogonal generalised Wannier functions
(NGWFs)\cite{SKYLARIS2002-1} and the
matrix $\mathbf{K}$ is called the density kernel\cite{MCWEENY1}.
$\mathbf{K}$ can be made sparse by enforcing the condition
$K^{\alpha \beta} =0$ when  $|\mathbf{R}_{\alpha}
-\mathbf{R}_{\beta}| > r_{\textrm{\scriptsize  cut}}$,
where $\mathbf{R}_{\alpha}$ and $\mathbf{R}_{\beta}$
are the centres of the localisation regions of NGWFs
$\phi_{\alpha}(\mathbf{r})$ and $\phi_{\beta}(\mathbf{r})$, respectively.

\begin{figure}

\centering

\scalebox{0.5}{\includegraphics*[3.0cm, 4.0
cm][25cm,20cm]{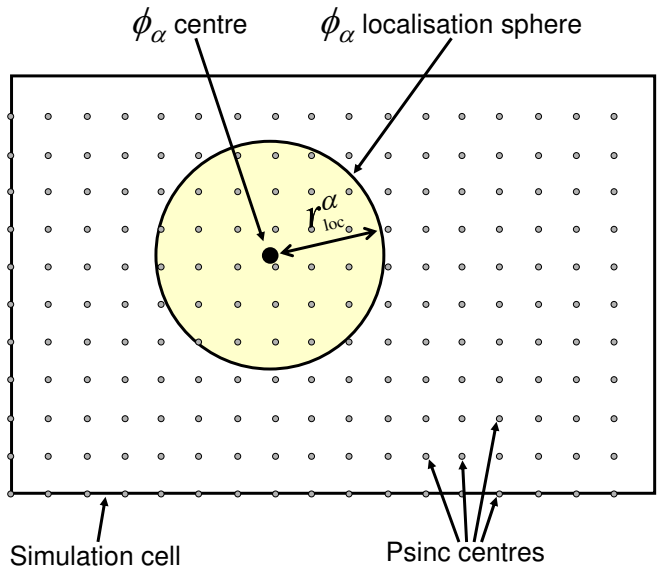} } \caption{Imposing
localisation on a NGWF ($\phi_{\alpha}$). The NGWF is expanded only
in the psinc functions whose centres fall inside its localisation
sphere. \label{ngwf-loc} }

\end{figure}

\textsc{onetep} belongs to the category of methods that aim for high
accuracy by optimising self-consistently the energy
 not only with respect to $\mathbf{K}$ but also with respect
 to the NGWFs
 \cite{GILLAN95,GILLAN97,HAYNES97, FATTEBERT2000,FATTEBERT2004,PASK2005}.
 In \textsc{onetep} the NGWFs are expanded in a basis set of
periodic cardinal sine (psinc) functions
\cite{SKYLARIS2002-1,MOSTOFI2003}, also known as Dirichlet or
Fourier Lagrange-mesh functions\cite{BAYE1986,VARGA2004}. Each psinc
function is centred on a particular point of a regular real-space
grid. Figure \ref{ngwf-loc} shows how this property is used to
impose localisation on the NGWFs within predefined spherical
regions.

\section{Basis set convergence} \label{convergence-section}

Since the computational cost of a DFT calculation increases with the
size of the basis set it is important to be able to converge
calculated properties to the desired accuracy using the smallest
possible basis set. The most convenient way to achieve this is to
improve the basis set systematically. For instance, the quality of a
plane wave basis \cite{PAYNE_REVIEW} is increased via a single
parameter, the kinetic energy cut-off. At the other end of the
spectrum are atomic orbital (AO) basis sets which do not span space
in a uniform manner and whose systematic refinement is not
straightforward. An AO basis is defined by a number of independent
features, such as the number of functions per atom, and their radial
and angular shapes. Furthermore, unlike plane waves, AO basis sets
are not orthogonal and consequently the undesired effect of linear
dependence can often hinder efforts to improve their quality.
Nevertheless, numerous careful attempts have been made to construct
series of atomic basis sets which demonstrate systematic improvement
to varying degrees
\cite{SANKEY1989,KENNY2000,JUNQUERA2001,ANGLADA2002}. Particular
attention has been paid to Gaussian\cite{GAUSSIANS} functions where
the series of even-tempered \cite{BARDO1974}  and
correlation-consistent \cite{cc-pVTZ1} basis sets are amongst the
most well-known cases of AO bases with systematic behaviour. In
\textsc{onetep} our psinc basis is constructed from plane waves in
such a way that it fully retains their desirable properties of
orthogonality and systematicity whilst being localised.

\begin{figure}

\centering

\scalebox{0.75}{\includegraphics*[3.0cm, 6.0
cm][25cm,18cm]{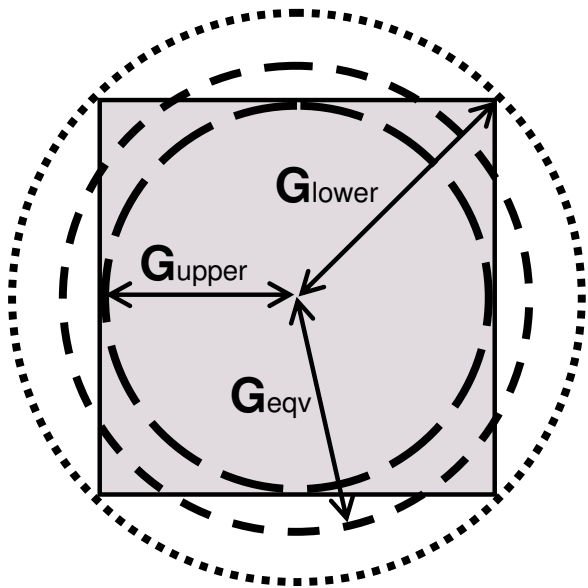} } \caption{The psinc basis of
\textsc{onetep} is constructed from a cube of wavevectors.
 Conventional plane wave approaches such as
\textsc{castep} define their plane wave basis from a sphere of
wave vectors. Three choices of such spheres that could be
used to compare \textsc{onetep} and \textsc{castep}
calculations are shown. \label{square-and-circle} }

\end{figure}

It is important to note that the set of plane waves which
constitute the psinc functions is different from that
in a typical plane wave calculation. The relation between the
two is clarified in Figure~\ref{square-and-circle}. The psinc basis
set is constructed from plane waves
$\mathrm{e}^{i\mathbf{G}\cdot\mathbf{r}}$
with wave-vectors $\mathbf{G}$ belonging to a cube of side-length
$2\mathbf{G_{\textrm{\scriptsize upper }} }$ in reciprocal space.
On the other hand, conventional plane wave approaches\cite{PAYNE_REVIEW}
 such as the \textsc{castep} code
 \cite{NEWCASTEP} construct their basis from a sphere of
$\mathbf{G}$ vectors. Therefore, to compare \textsc{onetep}
calculations with a code such as \textsc{castep}, we need to decide
first on the most appropriate sphere of wavevectors. Figure
\ref{square-and-circle} shows some choices for the radius of this
sphere: $\mathbf{G_{\textrm{\scriptsize upper}}}$ (sphere inscribed
in cube, the \textsc{castep} basis is a subset of the \textsc{onetep}
basis), $\mathbf{G_{\textrm{\scriptsize eqv}}}$ (sphere has
equal volume with cube, \textsc{onetep} and \textsc{castep}
basis sets have an equal number of functions with most of them
in common) and $\mathbf{G_{\textrm{\scriptsize
lower}}}$ (sphere circumscribes the cube, the \textsc{castep} basis
is a superset of the \textsc{onetep} basis).

\begin{figure}

\centering

\scalebox{0.40}{\includegraphics*[3.0cm, 9.0
cm][25cm,16cm]{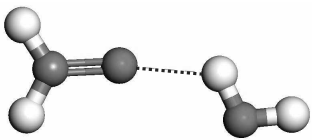} } \caption{The molecular structure of the
formaldehyde-water hydrogen bonded complex used in our tests
(not equilibrium geometry).
\label{h-bond}}

\end{figure}

In order to examine the strengths and weaknesses
of \textsc{onetep} compared to conventional plane wave and AO
approaches we have carried out a series of tests on the hydrogen bond
in the formaldehyde-water complex shown in Figure \ref{h-bond}. This is a
 rather sensitive test as hydrogen bonds
are amongst the weakest and longest chemical bonds, yet they are
very important because they are commonly encountered as major
contributors to the structural stability and function of most
biological macromolecules such as proteins, DNA and sugars
\cite{BRANDEN-TOOZE}. For the purpose of comparison we have used the
local density approximation (LDA)\cite{CEPERLEY1980, PERDEW1981}
exchange correlation (XC) functional.

Table \ref{castep-hbond-table} shows the binding energies we
obtained from calculations with \textsc{castep} for the three
kinetic energy cut-offs in Figure~\ref{square-and-circle}. Also
shown is the total energy of the bound complex. The core electrons
in these calculations were replaced by norm-conserving
pseudopotentials\cite{HAMANN1979,LEE1994,MILMAN1996}. Periodic
boundary conditions were used and the molecule was placed in a very
large cubic simulation cell (30 {\AA} $\times$ 30 {\AA} $\times$ 30
{\AA}) to ensure that the supercell approximation
\cite{PAYNE_REVIEW} holds extremely well.

The corresponding \textsc{onetep} results are shown in Table
\ref{onetep-hbond-table} for the same periodic simulation cell,
pseudopotentials and LDA XC functional. A total of 16 NGWFs were
used for the hydrogen bonded complex, one on each H atom and four on
each C and O atom. We have performed calculations for a wide range
of NGWF localisation sphere radii $r_{\textrm{\scriptsize loc}}$ and
we observe that the binding energy agrees with the converged
\textsc{castep} value\footnote{One kcal/mol is equal to 43.36 meV.}
to 1 meV, for $r_{\textrm{\scriptsize loc}}$ as small as 3.7 {\AA}.
The total energy  converges rapidly from above as a function of
$r_{\textrm{\scriptsize loc}}$, as expected for a basis set
variational method \cite{SKYLARIS2002-2}. We note that, once we are
converged with respect to $r_{\textrm{\scriptsize loc}}$, the
\textsc{onetep} result lies between the \textsc{castep} bounds shown
in Table~\ref{castep-hbond-table} and, as one would expect, agrees
closely with the 935 eV cut-off result
($\mathbf{G_{\textrm{\scriptsize eqv}}}$ sphere in
Figure~\ref{square-and-circle}). From here on we define the psinc
kinetic energy cut-off to be the kinetic energy cut-off of the plane
wave sphere with the same volume as the cube of our psinc basis.

Table~\ref{onetep-hbond-table} also shows the number of
self-consistency iterations taken to converge the total energy, and
we make the observation that this is independent of the localisation
region radius $r_{\textrm{\scriptsize loc}}$, which demonstrates
that our method does not suffer from the ``superposition
ill-conditioning'' described in Ref.~\onlinecite{BOWLER1998}.

\begin{table}
\caption{Calculations on the formaldehyde-water complex with
\textsc{castep} \cite{NEWCASTEP}. \label{castep-hbond-table}}
\begin{tabular}{l c c} \hline \hline
Kinetic energy                    &  Total energy  &  Binding energy \\
\multicolumn{1}{c}{cut-off (eV)}  &    (eV/atom)   &      (meV)
\\ \hline
608  ($\propto \mathbf{G_{\textrm{\scriptsize upper}}^2}$)   &   -154.444   &  145  \\
935  ($\propto \mathbf{G_{\textrm{\scriptsize eqv}}^2}$)     &   -155.044   &  149  \\
1823 ($\propto \mathbf{G_{\textrm{\scriptsize lower}}^2}$)   &   -155.082   &  148  \\
\hline \hline
\end{tabular}
\end{table}
\begin{table}
\caption{Calculations on the formaldehyde-water complex with
\textsc{onetep}\cite{SKYLARIS2005}.  \label{onetep-hbond-table}}
\begin{tabular}{c c c c}  \hline \hline
$r_{\textrm{\scriptsize  loc}}$ (\AA) & Number of  & Total energy & Binding energy \\
                                      & iterations &   (eV/atom)  & (meV) \\ \hline
 2.6 &  13 &  -154.789    &   168         \\
 3.2 &  13 &  -154.890    &   155         \\
 3.7 &  11 &  -154.914    &   150         \\
 4.2 &  11 &  -154.921    &   148         \\
 4.8 &  12 &  -154.924    &   148         \\
\hline\hline
\end{tabular}
\end{table}
\begin{table}
\caption{ Calculations on the formaldehyde-water complex with NWChem
\cite{NWCHEM4.5} using Gaussian basis sets of increasing size.
\label{nwchem-hbond-table}}
\begin{tabular}{l c c c}  \hline \hline
Basis name & Number & Binding energy & Counterpoise-corrected        \\
           & of AOs &     (meV)      & binding energy (meV)  \\ \hline
 STO-3G  \cite{STO3G1}                              &  19  &  91   & 39 \\
 3-21G \cite{BINKLEY1980}                           &  35  &  186  & 92   \\
 6-31G \cite{HEHRE1972}                             &  35  &  171  & 128  \\
 6-31+G* \cite{6-31+Gs-1, 6-31+Gs-2}                &  65  &  159  & 143  \\
 6-31++G** \cite{6-31+Gs-1, 6-31+Gs-2}              &  81  &  162  & 147  \\
 cc-pVDZ \& diffuse \cite{DUNNING1989,KENDALL1992}  & 111  &  153  & 146  \\
 cc-pVTZ  \cite{DUNNING1989}                        & 165  &  157  & 133  \\
 cc-pVTZ \& diffuse \cite{DUNNING1989,KENDALL1992}  & 265  &  149  & 147   \\
 cc-pVQZ  \cite{DUNNING1989}                        & 350  &  151  & 140 \\
 cc-pVQZ \& diffuse \cite{DUNNING1989,KENDALL1992}  & 535  &  148  & 147   \\  \hline\hline
\end{tabular}
\end{table}

To complete our comparison we present in Table
\ref{nwchem-hbond-table} calculations with the AO approach as
implemented in the NWChem\cite{NWCHEM4.5} quantum chemistry program
which uses Gaussian basis sets and a closely related formula
\cite{CEPERLEY1980,VWN} for the LDA XC functional. In this approach
the core electrons are treated explicitly and the molecules are
virtually isolated in space as the calculations are done with open
boundary conditions. The total number of AOs (contracted Gaussian
functions) for the whole formaldehyde-water complex for each basis
set is also shown in Table \ref{nwchem-hbond-table}.

From Table \ref{nwchem-hbond-table} we observe that the convergence
of the total energy of the complex is neither uniform nor rapid, as
a consequence of the fact that the different features, e.g., diffuse
functions etc., introduced to the basis set affect the energy to
different extents. We also note that the size of the basis set
 required to reach the same
level of accuracy as \textsc{onetep} is very large. The calculations
with the Gaussian basis set suffer from basis set superposition
error (BSSE) and thus in Table \ref{nwchem-hbond-table} we also give
a column with binding energies calculated with the counterpoise
correction method of Boys and Bernardi \cite{BOYS1970}. This costly
correction procedure significantly improves the binding energies
obtained with the medium sized basis sets (6-31+G* and 6-31++G**).

\section{Nanostructures, crystals and biomolecules}
\label{results-section}

In this section we present several examples of calculations on
systems with around 1000 atoms. Materials and molecules with this
number of atoms are usually beyond the capabilities of conventional
cubic-scaling approaches.

\subsection{Nanostructures: Carbon nanotubes} \label{tubes-subsection}

Carbon nanotubes are at the centre of many nanotechnology
applications because of their unique electronic and mechanical
properties\cite{NANOTUBE-BOOK}. From a structural point of view
nanotubes are seamless cylinders of graphene, which can be either
semiconducting or metallic. A method such as \textsc{onetep} where
linear-scaling is achieved by taking advantage of the exponential
decay of the density matrix present in insulators is not expected to
be efficient on metallic systems where the decay is only
algebraic\cite{ISMAIL-BEIGI1999}. Metallic systems therefore present
a significant challenge and carbon nanotubes are an ideal test case
that can provide us with insight into how switching from a
non-metallic to a metallic system (while keeping all other factors
essentially unchanged) affects calculations where density matrix
truncation is applied. We have studied segments of metallic (10, 10)
armchair and semiconducting (20, 0) zig-zag carbon nanotubes.
\cite{NANOTUBE-BOOK, ODOM2000}

  The (10, 10) nanotube segments are constructed by
repeating identical units of 40 atoms while the (20, 0) segments
are made of
units of 80 atoms. For the (10, 10) nanotube we performed
\textsc{onetep} calculations on segments consisting of 8, 15, 16, 30
and 32 units ranging from 320 to 1280 atoms. For the (20, 0)
nanotube we used segments of 8 and 16 units with 640 and 1280
atoms respectively. As our
 nanotube segments were made of repeated identical
units we were able to perform with \textsc{castep} calculations
equivalent to \textsc{onetep} by using only a single unit
but equivalent Brillouin zone sampling.
\begin{figure}

\vspace{-8ex}

\centering

\scalebox{0.75}{\includegraphics*[3.3cm,
6cm][26cm,20cm]{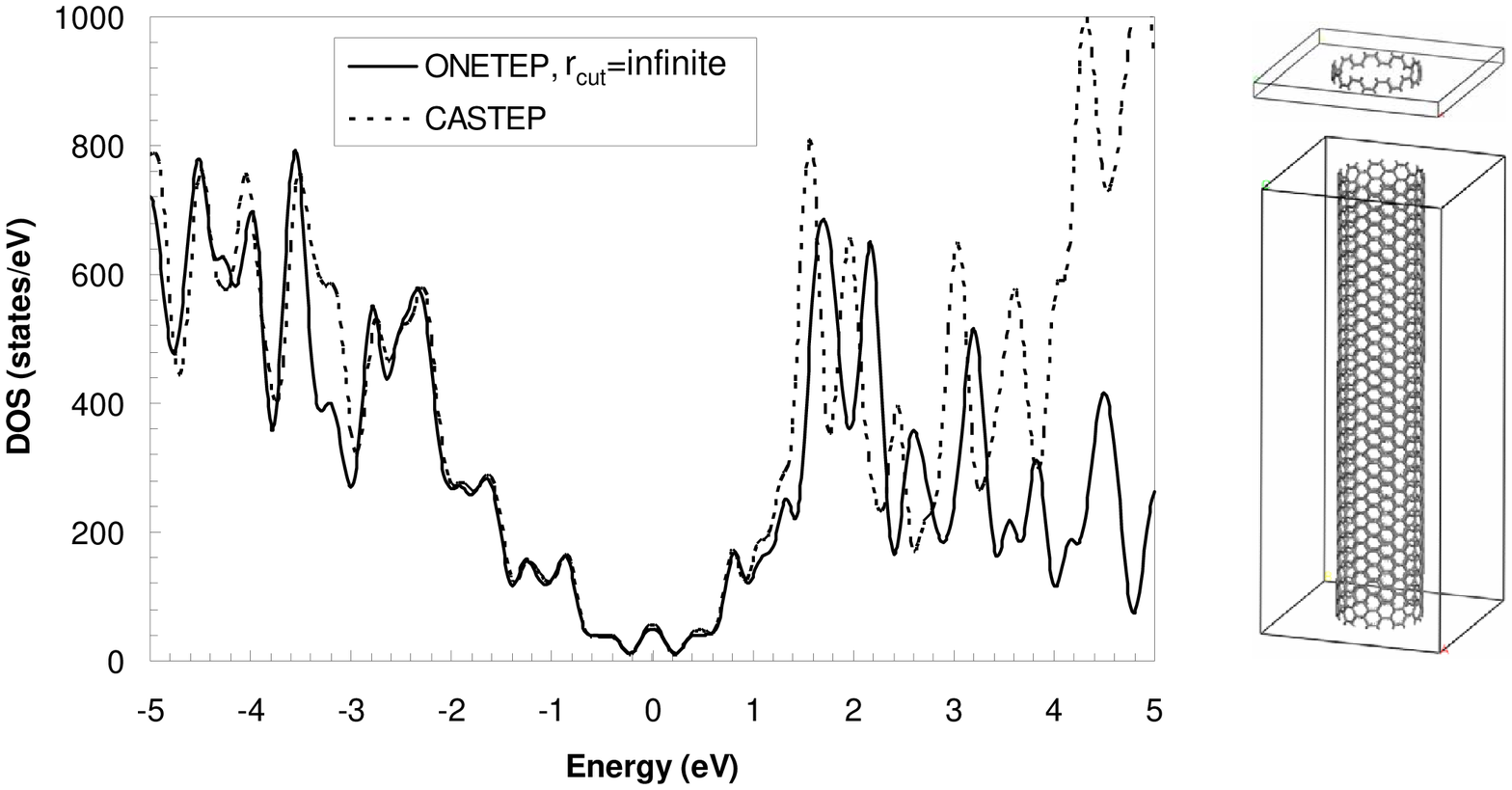} }

\scalebox{0.75}{\includegraphics*[3cm,
5cm][26cm,18cm]{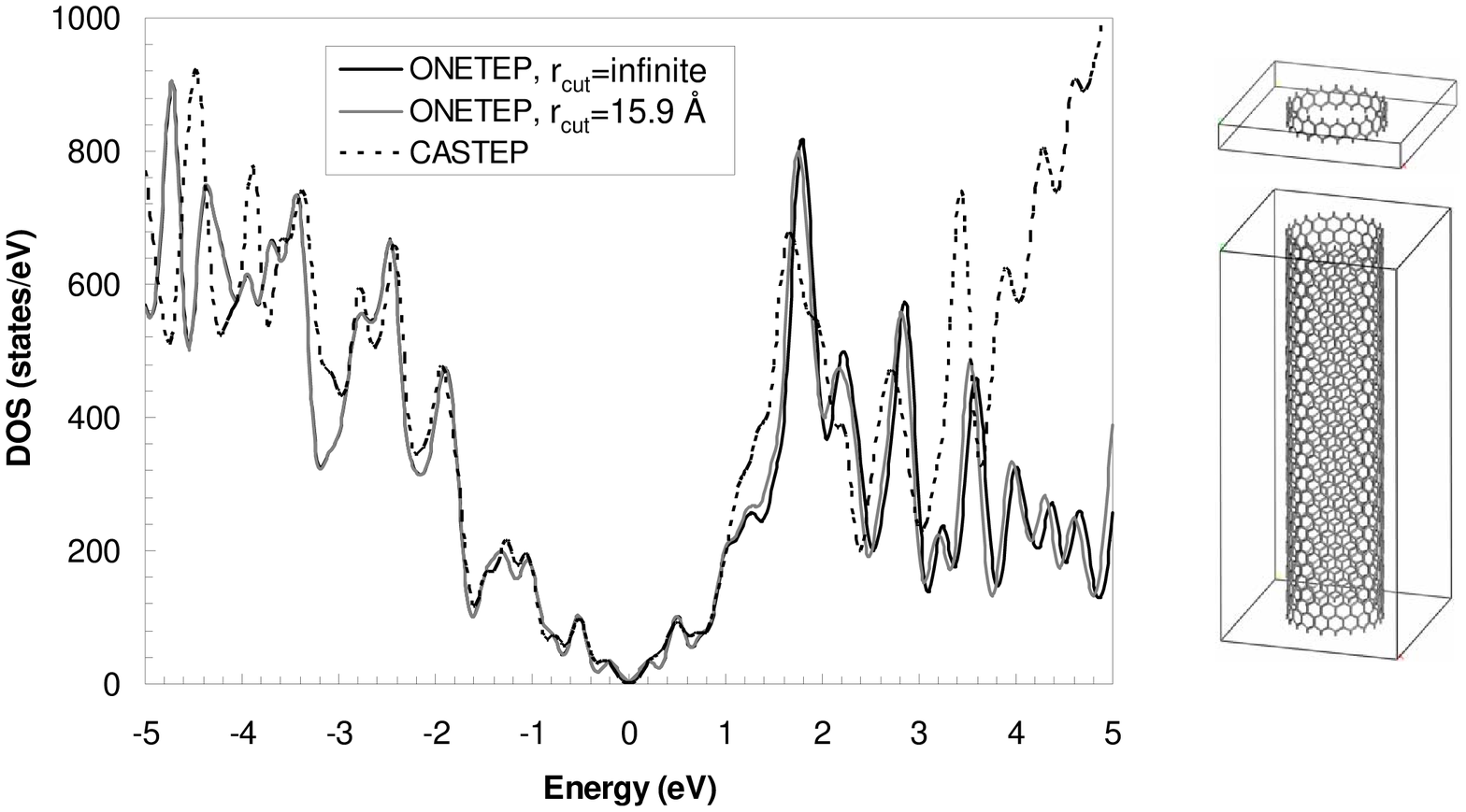} } \caption{Top panel: The
density of states (DOS) of a 30-unit segment of a (10, 10) metallic
nanotube as calculated with \textsc{onetep} and \textsc{castep}. On
the right the \textsc{onetep}
(30{\AA}$\times$30{\AA}$\times$73.30{\AA}) and \textsc{castep}
(30{\AA}$\times$30{\AA}$\times$2.44{\AA}) simulation cells are
shown. Bottom panel: The density of states (DOS) of a 16-unit
segment of a (20, 0) semiconducting nanotube as calculated with
\textsc{onetep} and \textsc{castep}. On the right the
\textsc{onetep} (30{\AA}$\times$30{\AA}$\times$67.78{\AA}) and
\textsc{castep} (30{\AA}$\times$30{\AA}$\times$4.24{\AA}) simulation
cells are shown.
 \label{nanotube-doses}}

\end{figure}
The same LDA \cite{CEPERLEY1980, PERDEW1981}
XC functional and
pseudopotential were used by both codes. The plane wave kinetic
energy cut-off of \textsc{castep} was set to 410 eV as was
 the  psinc kinetic energy cut-off of
\textsc{onetep}. In the \textsc{onetep} calculations the radii
$r_{\textrm{\scriptsize loc}}$  of the carbon NGWF localisation
spheres were 3.3 {\AA}. The nanotube segments were placed in
orthorhombic simulation cells with their axis aligned with the
z-axis. The dimensions of the cells along the x- and y-axes were 30
{\AA} $\times$ 30 {\AA}. These simulation cells ensured negligible
interaction of the nanotubes with their periodic images as the
diameter of the (10, 10) tubes is just 13.6 {\AA} and that of the
(20, 0) tubes is 15.6 {\AA}. In order to perform a detailed
comparison of the results between the two codes we diagonalised the
converged \textsc{onetep} Hamiltonian in the NGWF representation and
obtained canonical molecular orbitals. From these we constructed the
density of states (DOS) by smearing with Gaussians with a halfwidth
of 0.1 eV. Our results are shown in Figure \ref{nanotube-doses}. The
two codes give virtually identical DOS in the important regions of
1~eV below and above the Fermi level and very close agreement in the
region below -1 eV. In the region above 1 eV the agreement
deteriorates rapidly. This is not surprising as the NGWFs of
\textsc{onetep} are specifically optimised to describe the density
matrix which is composed of occupied bands and no emphasis is placed
on the description of the conduction bands. It is still remarkable
that the low-lying conduction band DOS is calculated correctly with
\textsc{onetep}.

As we make our (10, 10) nanotube segments longer, we increase the
number of closely-spaced \textbf{k}-points from the metallic band
structure of the nanotube that we fold into our equivalent $\Gamma$
point description of the band structure and the density matrix. We
have found that as the number of segments increases, it becomes more
and more difficult to impose a finite density kernel cut-off
threshold $r_{\textrm{\scriptsize cut}}$   in \textsc{onetep} while
maintaining any degree of accuracy. With the 30 and 32 unit segments
an infinite $r_{\textrm{\scriptsize cut}}$ becomes essential in
order to obtain useful results. In contrast, the (20, 0) nanotube
remains amenable to density kernel truncation as the length of its
segments is increased. For example in Figure \ref{nanotube-doses} we
show the DOS for the 16 unit segment generated with
$r_{\textrm{\scriptsize cut}}=\infty$
 and with $r_{\textrm{\scriptsize cut}}=15.9$ {\AA}
 and the
two curves essentially coincide. Our observations are thus
consistent with expected behaviour regarding the decay of the
density matrix in metallic and non-metallic systems at zero
temperature.

 \textsc{onetep} calculations with
 $r_{\textrm{\scriptsize cut}}=\infty$, while not
 linear-scaling, are still perfectly feasible.
 In particular,
 most computationally intensive steps
 such as the construction of the Hamiltonian matrix in the
 NGWF representation, the construction of the electronic
 charge density and the calculation of the derivatives
 of the NGWFs with respect to the expansion coefficients
 in the psinc basis depend only in the NGWF localisation
 sphere radii $r_{\textrm{\scriptsize loc}}$
  and are always perfectly
 linear-scaling independently of the value of $r_{\textrm{\scriptsize cut}}$.
 The only step that stops being linear-scaling when the
 density kernel $\mathbf{K}$ is no longer sparse
 is the optimisation of $\mathbf{K}$
 which is carried out by using variants of
 the Li-Nunes-Vanderbilt \cite{LI1} method and
 Haynes \cite{HAYNES99} penalty functional
 method which involve matrix multiplications.

It is also worth noting that unlike conventional plane wave
approaches where the memory and computation grows with the entire
volume of the simulation cell without distinction between vacuum and
atomic regions, \textsc{onetep} uses
algoritmns\cite{KINETIC_PAPER,TOTALENERGY_PAPER} which avoid
computation and storage in vacuum regions making thus possible
calculations in very large simulation cells as in this section.

\subsection{Solids: crystalline silicon}

\begin{figure}
\centering \scalebox{1.0}{\includegraphics*[3.3cm,
7.0cm][17cm,13cm]{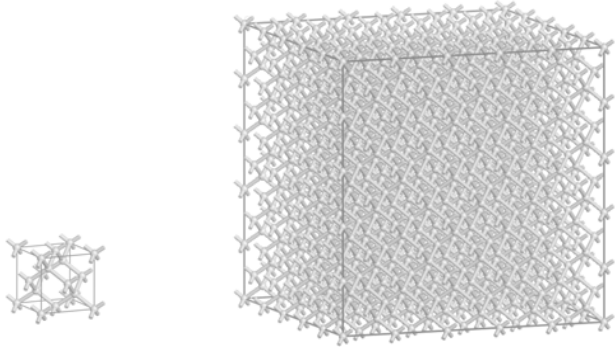} } \caption{Periodic
crystalline silicon. Left: The 8-atom cubic simulation cell used in
the calculations with \textsc{castep}. Right: The 1000-atom cubic
simulation cell used in the calculations with \textsc{onetep}.}
\label{cells-figure}
\end{figure}

Here we examine properties of pure crystalline silicon as calculated
by \textsc{onetep} and \textsc{castep}. For these calculations we
have used the LDA with a norm-conserving pseudopotential and plane
wave and psinc kinetic energy cut-offs of 283 eV. A cubic unit cell
of 1000 atoms was used in the \textsc{onetep} calculations and a
cubic unit cell of 8 atoms was used in the \textsc{castep}
calculations, with an equivalent 5$\times$5$\times$5
\textbf{k}-point mesh. The two cells are shown in Figure
\ref{cells-figure}.
\begin{figure}
\centering \scalebox{0.55}{\includegraphics*[0cm,
1cm][26cm,19cm]{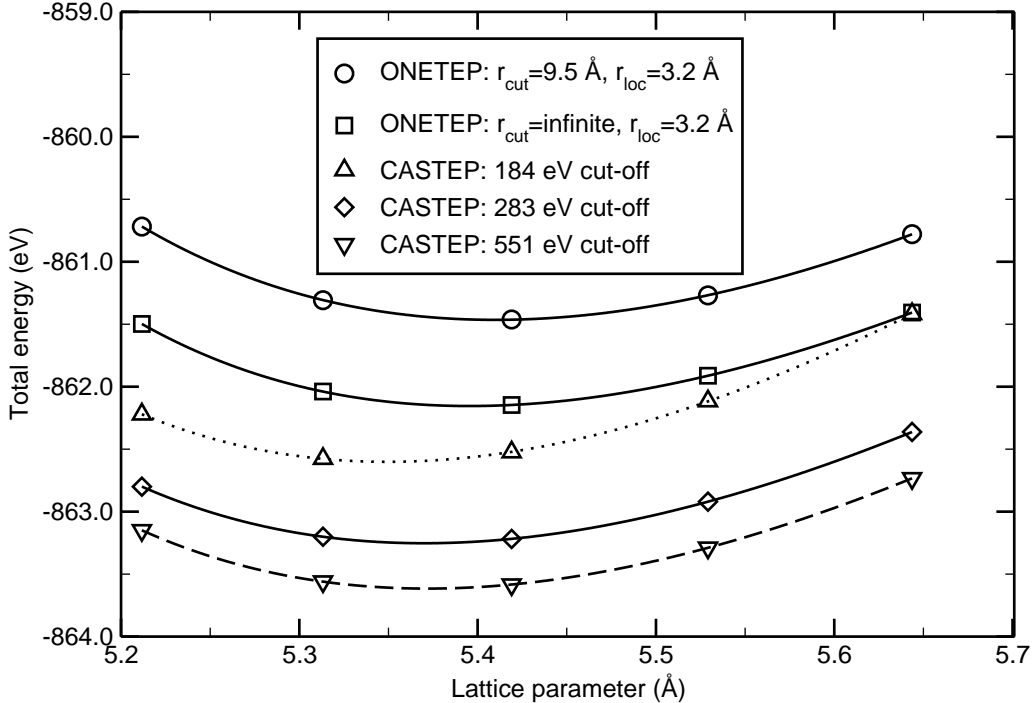} } \caption{The total energy per
8-atom unit cell of silicon as a function of the lattice parameter
for calculations with \textsc{castep} and \textsc{onetep}.}
\label{lattice-constants-figure}
\end{figure}
\begin{table}
\caption{Lattice constant and bulk modulus of perfect crystalline
silicon as calculated by \textsc{castep}, \textsc{onetep} and
experiment.} \label{silicon-table}
\begin{tabular}{lccc} \hline \hline
Method & Kinetic energy & Lattice constant  & Bulk modulus  \\
       & cut-off (eV)   &    (\AA)          &    (GPa)       \\\hline
\multirow{3}{*}{\textsc{castep}, constant E$_{\textrm{\scriptsize cut}}$}
& 184 &  5.410   &   94.7  \\
& 283 &  5.392   &   94.4  \\
& 551 &  5.383   &   95.9  \\
\hline
\multirow{3}{*}{\textsc{castep}, constant $N_{\textrm{\scriptsize PW}}$}
& 184 &  5.359   &  109.1  \\
& 283 &  5.380   &   96.1  \\
& 551 &  5.382   &   96.6  \\
\hline \textsc{onetep}, constant $N_{\textrm{\scriptsize psinc}}$,
$r_{\textrm{\scriptsize cut } } = \infty$
& 283 &  5.406   &   99.6  \\
\textsc{onetep}, constant $N_{\textrm{\scriptsize psinc}}$,
$r_{\textrm{\scriptsize cut } } = 9.5$ {\AA}
& 283 & 5.421 & 99.5 \\
\hline
Experiment  & &  5.430   &  100.0 \\
\hline\hline
\end{tabular}

\end{table}
We should note that in a code like \textsc{castep} there are two
ways to define the basis set while varying the energy with respect
to the
 lattice parameter. One can either keep the kinetic energy cut-off
E$_{\textrm{\scriptsize cut}}$ constant or keep the number of plane
wave basis functions $N_{\textrm{\scriptsize PW}}$ constant. The
latter approach is conceptually closer to the way the
\textsc{onetep} calculations are performed in these cases as it is
the number of psinc functions that is kept constant, which is
equivalent to keeping constant the number of plane waves in the cube
of Figure \ref{square-and-circle}. Furthermore, in the
\textsc{onetep} calculations, when varying the lattice parameter, it
is important to scale $r_{\textrm{\scriptsize loc } }$ and $r_{\textrm{\scriptsize
cut} }$ proportionately.
Throughout this
section the values we report for these quantities correspond to a
lattice parameter of 5.43 {\AA}.

In Figure \ref{lattice-constants-figure} we show \textsc{castep}
constant-$N_{\textrm{\scriptsize PW}}$  plots of the total energy
per 8-atom cell as a function of the lattice parameter for kinetic
energy cut-offs which correspond to the ``upper bound" (184 eV),
``equivalent" (283 eV) and
 ``lower bound" (551 eV) cases of Figure \ref{square-and-circle}.
The two \textsc{onetep} curves
lie higher in energy than the \textsc{castep} curves because the
NGWF radii we used were only 3.2 {\AA} and the total energy is not
yet completely converged with respect to them. Nevertheless, the
physical properties that we calculate are already converged to a very
satisfactory level.

By fitting the calculated energies as a function of the lattice
parameter to the Birch-Murnaghan equation of state
\cite{MURNAGHAN1944} we obtained values for the lattice constants
and bulk moduli of crystalline silicon which we show in Table
\ref{silicon-table}. There is excellent agreement between the
\textsc{onetep} and \textsc{castep} constant-$N_{\textrm{\scriptsize
PW}}$ results at 283 eV. For the case of the infinite
$r_{\textrm{\scriptsize cut}}$ the lattice constants agree to 0.5\%
and the bulk moduli to 3.6\% while for the case of the 9.5 {\AA}
$r_{\textrm{\scriptsize cut}}$ the lattice constants agree to 0.8\%
and bulk moduli to 3.5\%.

The bulk modulus is a quantity which is sensitive to
calculation parameters and difficult to converge. Even
between the \textsc{castep} calculations with the highest cut-off of
551 eV there remains a difference of 0.7\% between the  bulk modulus
values obtained with constant E$_{\textrm{\scriptsize cut}}$ and
constant $N_{\textrm{\scriptsize PW}}$ while the lattice constant
difference in this case is reduced to only 0.02\%.

\subsection{Biomolecules: breast cancer susceptibility proteins}

Biomolecules are generaly too large for conventional DFT
calculations. Nevertheless a number of insightful studies have been
carried out where a small fragment can be isolated from the rest of
the biomolecule \cite{MOLTENI1999,SEGALL2002}. Obviously this
approach cannot be applied in cases where the interactions extend
over a large area, e.g., the case of two large proteins bound to
each other. \textsc{onetep} can offer great advantages in the study
of such molecules since it allows one to perform calculations either
on entire biomolecules or at least on segments large enough to
contain the entire area of interaction. An example of the latter
case is the RAD51-BRCA2 protein complex for which we present
preliminary results in this section.

The breast cancer susceptibility protein \cite{VENKITARAMAN2002}
BRCA2 regulates the function of RAD51, an enzyme involved in DNA
recombination. Crucial to this process is the specific interaction
between RAD51 and a BRC motif (sub-region) in BRCA2. There are eight
slightly different versions of the BRC motif in a single BRCA2
protein and each of these motifs can interact with a RAD51 protein.
Recently the structure of RAD51 bound to one of the BRC motifs
(BRC4) has been elucidated by high resolution X-ray diffraction,
revealing in a qualitative manner the nature of the interactions at
the site of contact between the two proteins \cite{PELLEGRINI2002}.
Amino acids with both polar and hydrophobic side chains are involved
in these interactions. With this crystal structure as our starting
point, we have used calculations with \textsc{onetep} to predict the
strength of the binding between the two proteins.
\begin{figure}
\centering \scalebox{0.75}{\includegraphics*[3cm,
9cm][26cm,16cm]{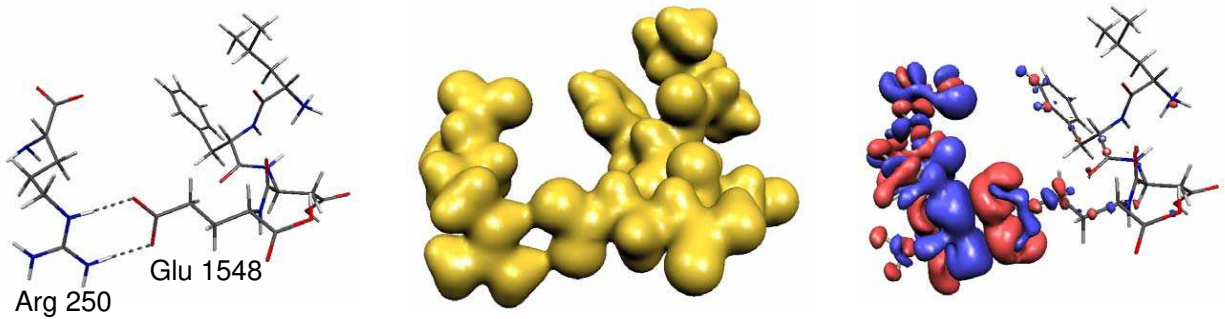} } \caption{97-atom segment
which includes the bonding interactions between the Arg 250 - Glu
1548 residues of the BRCA2-RAD51 complex. Left: stick model of the
atomic structure. Middle: isosurface of the electronic change
density at a value of 0.02 $e^-/a_0^{3}$ from the \textsc{onetep}
calculation. Right: isosurface of the electronic charge density
difference due to bonding at a value of 0.00075 $e^-/a_0^{3}$ from
the \textsc{onetep} calculation.} \label{arg1-figure}
\end{figure}
\begin{figure}
\centering
 \scalebox{0.75}{\includegraphics*[2.7cm,
7.5cm][26.5cm,17cm]{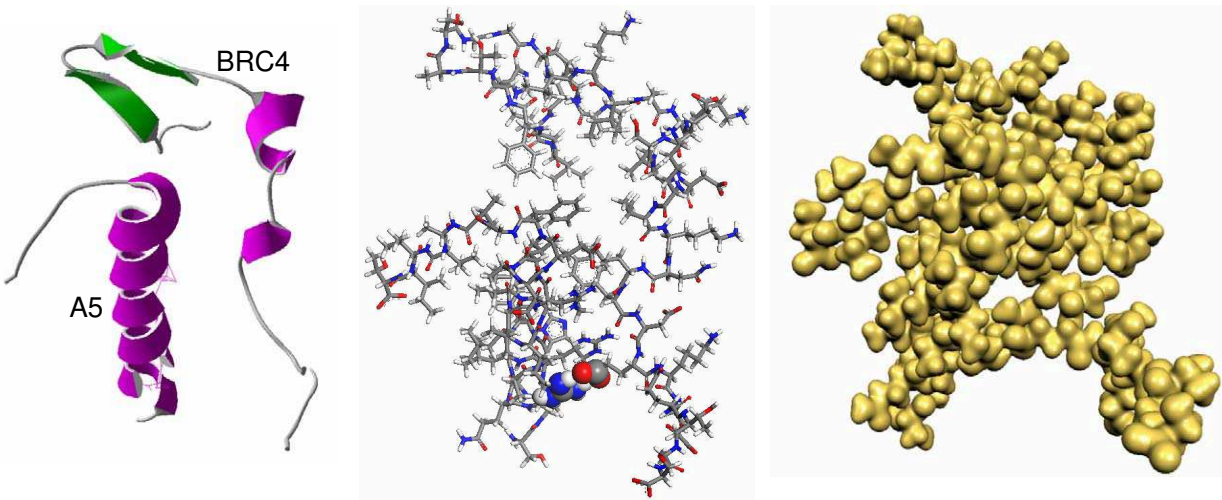} } \caption{The 988-atom A5-BRC4
complex. Left: tertiary structure. Middle: stick model in atomic
detail with the side groups of Arg 250 and Glu 1548 shown in
space-filling form. Right: isosurface of the electronic charge
density at a value of 0.02 $e^-/a_0^{3}$ from the \textsc{onetep}
calculation. } \label{bA-figure}
\end{figure}
The 988-atom protein segment we have studied here (Figure
\ref{bA-figure}) consists of the entire BRC4 motif and only the A5
$\alpha$-helix of the RAD51. According to Pellegrini et al.
\cite{PELLEGRINI2002} the major bonding interaction between A5 and
BRC4 is a polar interaction involving hydrogen bonding between the
side chains of arginine 250 of A5 and glutamic acid 1548 of BRC4. We
have first studied just this interaction in isolation by cutting a
very small 97-atom segment from the crystal structure of the
proteins (Figure \ref{arg1-figure}) which contains the relevant
amino acids Arg 205 and Glu 1548. The two hydrogen bonds between the
cationic side chain of the arginine and the anionic chain of the
glutamic acid of the segment are depicted in Figure
\ref{arg1-figure}. For this segment we were able to perform
calculations with both \textsc{onetep} and \textsc{castep}. We have
used norm-conserving pseudopotentials, plane wave and psinc kinetic
energy cutoffs of 608 eV, cubic simulation cells of 25 {\AA}
$\times$ 25 {\AA} $\times$ 25 {\AA} and the Perdew-Burke-Ernzerhof
(PBE) \cite{PERDEW1996} exchange-correlation functional. The radii
$r_{\textrm{\scriptsize loc } }$ of the NGWF localisation spheres
were set to 3.2 {\AA} for the hydrogen atoms and 3.6 {\AA} for all
other atoms. The binding energies between these two fragments as
calculated by \textsc{castep} and \textsc{onetep} are 2.78~eV and
2.79~eV respectively. Besides the excellent agreement between the
two codes it is worth observing that this binding energy is large in
comparison to the energy of two regular hydrogen bonds (which
individually range from about 100 to 300~meV). It appears that the
bulk of the binding strength comes from the electrostatic
interaction between the +1 charge of the arginine side group and the
-1 charge of the carboxyl of the glutamic acid. Indeed, the
classical electrostatic energy of a system of two point charges of
+1 and -1 atomic units separated by the same distance as the centres
of the side chains of these amino acids, is about 3.10~eV which is
rather close to the calculated binding energy. Calculations with the
NWChem code with the 6-31+G* Gaussian basis set and the PBE
functional produced a binding energy of 2.87~eV which after the
counterpoise correction for BSSE became 2.82~eV, in very good
agreement with \textsc{castep} and \textsc{onetep} given the level
of accuracy that can be reached with a Gaussian basis set of this
quality (section \ref{convergence-section}).

For the BRC4-A5 complex of Figure \ref{bA-figure} the same
calculation parameters as for the 97-atom segment were used except
for the orthorhombic 60 {\AA} $\times$ 50 {\AA} $\times$ 60 {\AA}
simulation cell. The binding energy between the whole A5 helix and
the BRC4 motif that we obtained from our calculations with
\textsc{onetep} is 5.67~eV. This is about twice as much as the
binding energy of the small segment of Figure \ref{arg1-figure} and
it shows that the remaining interactions between A5 and BRC4, though
small individually, cannot be neglected.

\section{Conclusions} \label{conclusions-section}
In comparison with two well-established cubic-scaling
density functional methods, we have demonstrated
that \textsc{onetep} can routinely achieve the highest levels of
accuracy that are possible with these methods. Amongst the factors
that make this possible is the fact that in \textsc{onetep} the calculated
properties converge rapidly with the radii of the localisation
spheres of nonorthogonal generalised Wannier functions (NGWFs) and
the rate of self-consistent convergence is affected neither by the
size of these regions nor the number of atoms. Next we have
demonstrated the wide applicability of the method by presenting
exploratory calculations in systems of about 1000 atoms from a wide
variety of materials. We have studied semiconducting and
metallic nanotubes, crystalline silicon,
and the complex of two bound proteins that can play a role in the
development of breast cancer. In all these cases we have managed to
obtain excellent agreement with \textsc{castep} in comparing either
on smaller systems of the same material or, where possible by the use
of \textbf{k}-points, in systems of equivalent size. These results
confirm that \textsc{onetep} is a robust, highly-accurate
linear-scaling density functional approach which makes possible a
whole new level of large scale simulation in systems of interest to
nanotechnology, biophysics and condensed matter physics.

\begin{acknowledgments}
C.-K.~S. would like to thank the Royal Society for a University
Research Fellowship. P.~D.~H. would like to thank Sidney Sussex
College Cambridge for a Research Fellowship. A.~A.~M. would like to
thank Christ's College Cambridge for a Research Fellowship. We are
indebted to Professor Ashok Venkitaraman for introducing us to the
BRCA2-RAD51 problem. C.-K.~S. would like to thank Pamela Schartau
for useful discussions on the theory of protein structure. The
computing resources of the Cambridge-Cranfield High Performance
Computing Facility have been essential for this work.
\end{acknowledgments}

\bibliography{../../../../BIBS/cav_base,../../../../BIBS/journal_base}

\end{document}